\documentclass{iau}
\usepackage{graphicx}

\title[Sgr A* in bright state] 
{Mini-spiral as source of material for Sgr A* in bright state}

\author[B.\ Czerny, V.\ Karas, D.\ Kunneriath, \& T.\ K.\ Das]   
{Bozena Czerny,$^1$ Vladim\'{\i}r Karas,$^2$ Devaky Kunneriath,$^2$\\[3pt] \and Tapas K. Das$^3$}

\affiliation{$^1$\,Copernicus Astronomical Center, Bartycka 18, 00-716 Warsaw, Poland  \\ email: {\tt bcz@camk.edu.pl } \\[\affilskip]
$^2$\,Astronomical Institute, Academy of Sciences, \\ Bo\v{c}n\'{\i} II 1401, CZ-14131 Prague, Czech Republic  \\ email: {\tt vladimir.karas@cuni.cz,  devaky@ig.cas.cz} \\[\affilskip]
$^3$\,Harish Chandra Research Institute, Allahabad 211019, India \\ email: {\tt tapas@mri.ernet.in}}

\pubyear{2012}
\volume{290}  
\jname{Feeding compact objects: Accretion on all scales}
\editors{C.\ M. Zhang, T. Belloni, M.\ M\'endez \& S.\ N.\ Zhang, eds.}

\begin{document}

\maketitle

\begin{abstract}
The question of the origin of the gas supplying the 
accretion process is pertinent especially in the context of enhanced activity of Galactic Center during 
the past few hundred years, seen now as echo from the surrounding molecular clouds,  and the currently 
observed new cloud approaching Sgr A*. We discuss the so-called Galactic Center mini-spiral as a possible 
source of material feeding the supermassive black hole on a 0.1 parsec scale. The collisions between
individual clumps reduce their angular momentum.  and set some of the clumps on a plunging trajectory.

We conclude that the amount 
of material contained in the mini-spiral is sufficient to sustain the luminosity of Sgr A* at the 
required level. The accretion episodes of relatively dense gas from the mini-spiral passing through a 
transient ring mode at  $\sim 10^4$ Rg  provide a viable scenario for the bright phase of Galactic Center. 
\keywords{Galaxy: center; Accretion, accretion disks; Black hole physics}
\end{abstract}

\firstsection 
\section{Introduction}

The supermassive black hole at the Milky Way center (Sgr A*) has the mass of about $4.4 \times 10^6 M_{\odot}$
(Genzel et al. 2010). It presently 
remains in a very quiet state. However, observations of
the X-ray reflection from molecular clouds seem to imply that
Sgr A* was orders of magnitude brighter only a few hundred
years ago (see Ponti et al. 2010, and further references therein). It is 
interesting to note the recent discovery of a cloud identified at the distance of only a 
few thousands gravitational radii; see Gillessen et al.\ (2012). The cloud moves along a highly eccentric
orbit, and so the latter authors suggest that will reach the 
minimum distance from the Sgr A* in 2013.  Here we consider the mini-spiral as a possible source
of the intermittently inflowing material.

\begin{figure}[t]
\begin{center}
\hfill~
 \includegraphics[width=0.38\textwidth]{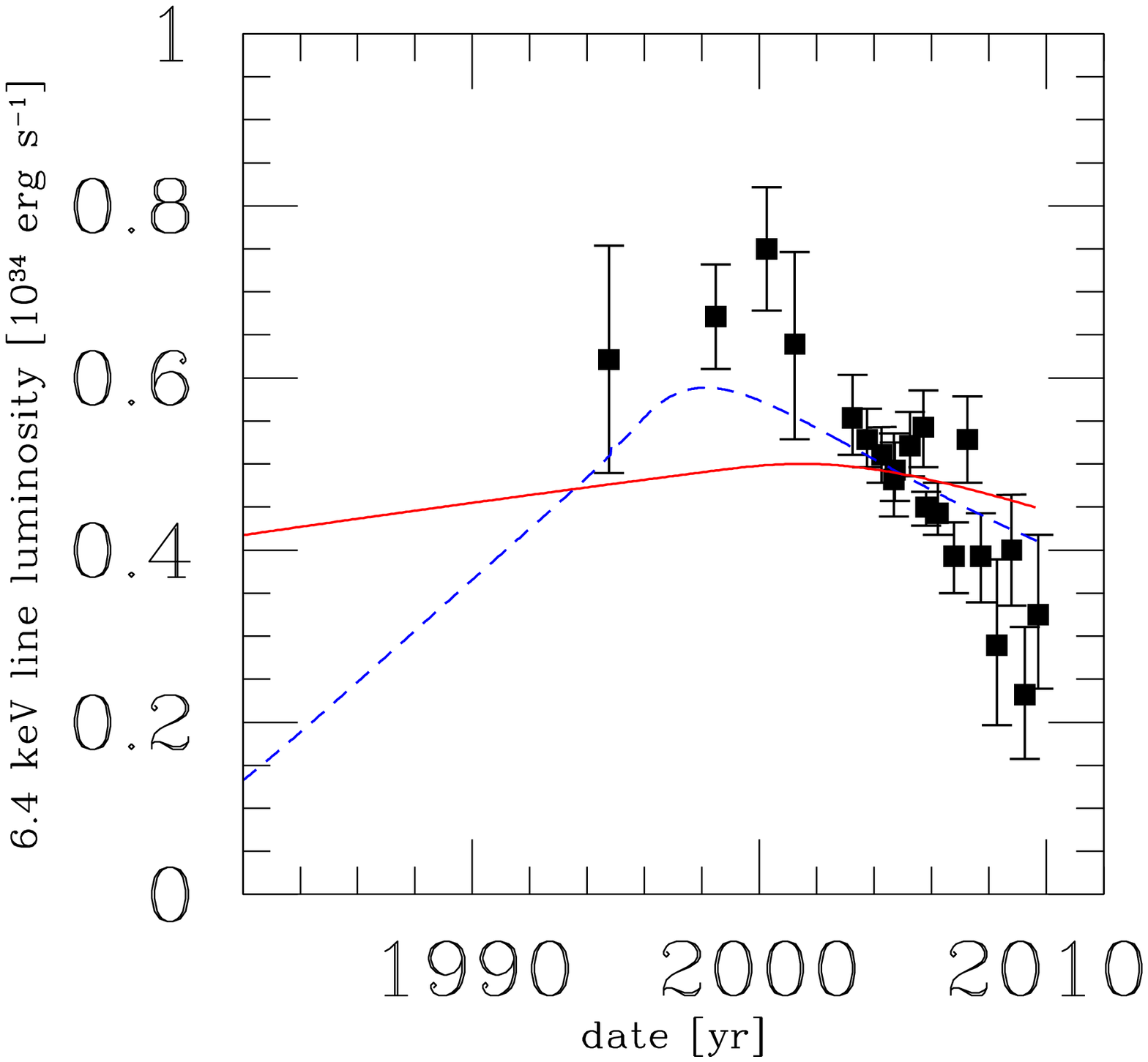}
\hfill~
 \includegraphics[width=0.38\textwidth]{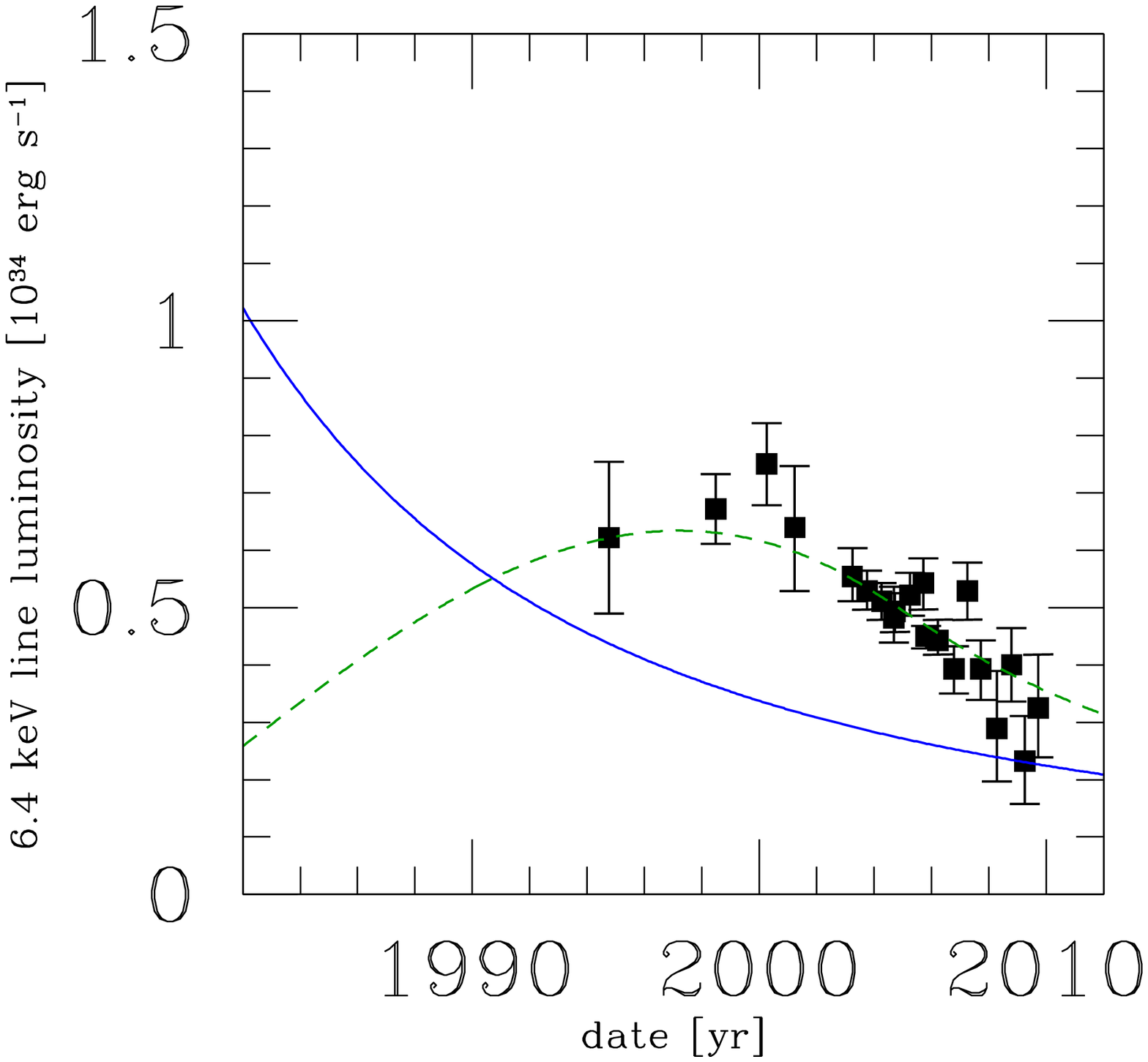}
\hfill~
 \caption{Two examples of model fits that are consistent with data points from Terrier et al. (2010),
 Yu et al. (2011). Left: events without reprocessing (single event: $t_{\rm event} = 1990.7$, $t_{\rm decay} = 12$ 
 years, dashed blue; a series of events  $t_{\rm event} = 1960$--$1995.5$, $t_{\rm decay} = 9$ years). 
 Right:  single flare event with a decay time-scale $t_{\rm decay}=9$ 
years (blue continuous line -- intrinsic profile ($t_{\rm{}event}=1969$),
green dashed line -- the scattered signal assuming the cloud size 5~pc.}
\end{center}
\end{figure}

\section{Model}

The Galactic Center black hole is surrounded by the minispiral, a complex system of gas and dust
(Zhao et al. 2010, Kunneriath et al. 2012). The currently adopted explanation assumes a superposition
of three elongated clumpy streams of material in a
roughly Keplerian motion. The Northern Arm and the Eastern Arm appear to collide
about 0.1--0.2 pc behind Sgr A* (Zhao et al. 2010). 
The hot gas can accrete directly onto the black hole.  The Bondi radius for a cold
gas is much smaller, so some kind of dissipative losses of angular momentum in collisions are needed.

The next stage of the evolution proceeds by 
action of viscosity. The time evolution of a gas located initially in a ring can be described analytically, 
and the accretion rate at the inner edge of the newly formed disk is described by
the formula (Zdziarski et al. 2009)
\begin{equation}
\dot M(\tau) = {M_{\rm ring} \over 4 \pi R_{\rm circ}^2\, t_{\rm visc}} {(2 \mu^2)^\mu \over \Gamma(\mu)}\, \tau^{-1 -\mu}\, \exp\left(-{2 \mu^2 \over \tau}\right),
\end{equation}
where $\mu$ describes the viscosity, and $\tau_{\rm visc}$ depends on the circularization 
radius.
We adopt $\mu=2/3$ in further considerations. 
We model the decay timescale of the K$\alpha$ reflection from the molecular cloud B2 
assuming a range of accretion events distributed in time and/or in angular momentum. We also
consider the smearing of the signal by the reprocessing due to a finite size of the cloud.

\section{Results}
We considered several examples of the model lightcurves, with and without the effect of reprocessing by the 
extended region of the B2 molecular cloud. 
Fig.~1 left panel shows the evolution for the case of a narrow accretion ring, while the right panel shows the result
for a whole range of angular momentum. 
For a slowly cooling material, the fitted decay time-scale 12 years (without reprocessing) implies an accretion event settling as a ring 
 $\sim 1.4 \times 10^4 R_{\rm g}$, i.e. only by factor 20 less 
than the inner edge of the mini-spiral, and the infalling matter can retain as much as 20 per cent of 
the initial angular momentum. Reprocessing slightly decreases this value to 9 years and introduces a considerable delay of the event.
For further details see our paper (A\&A submitted).

\end{document}